\documentclass[letterpaper,twocolumn,prl,
aps,showpacs,superscriptaddress,floatfix]{revtex4-2}

\usepackage[latin1]{inputenc}
\usepackage{bbm}
\usepackage{bm}
\usepackage[usenames]{color}
\usepackage{multirow}
\usepackage{amssymb}
\usepackage{amsbsy}
\usepackage{mathtools}
\usepackage{amsmath}
\usepackage{stmaryrd}
\usepackage{graphicx}
\usepackage{epsfig}
\usepackage{placeins}
\usepackage{bbold}
\usepackage{braket}
\usepackage{blindtext}
\usepackage[colorlinks,linkcolor=blue,citecolor=blue,urlcolor=blue]{hyperref}

\usepackage{scalerel}
\usepackage{tikz}
\usetikzlibrary{calc}
\usetikzlibrary{patterns}
\usetikzlibrary{svg.path}
\definecolor{orcidlogocol}{HTML}{A6CE39}
\tikzset{
  orcidlogo/.pic={
    \fill[orcidlogocol] svg{M256,128c0,70.7-57.3,128-128,128C57.3,256,0,198.7,0,128C0,57.3,57.3,0,128,0C198.7,0,256,57.3,256,128z};
    \fill[white] svg{M86.3,186.2H70.9V79.1h15.4v48.4V186.2z}
                 svg{M108.9,79.1h41.6c39.6,0,57,28.3,57,53.6c0,27.5-21.5,53.6-56.8,53.6h-41.8V79.1z M124.3,172.4h24.5c34.9,0,42.9-26.5,42.9-39.7c0-21.5-13.7-39.7-43.7-39.7h-23.7V172.4z}
                 svg{M88.7,56.8c0,5.5-4.5,10.1-10.1,10.1c-5.6,0-10.1-4.6-10.1-10.1c0-5.6,4.5-10.1,10.1-10.1C84.2,46.7,88.7,51.3,88.7,56.8z};
  }
}

\newcommand\orcid[1]{\!%
  \href{https://orcid.org/#1}{%
    \mbox{%
      \scaleto{%
        \begin{tikzpicture}[yscale=-1,transform shape]
          \pic{orcidlogo};
        \end{tikzpicture}
      }{8pt}%
    }%
  }%
}

\makeatletter

\begin{document}
\title{Diffusion constants from the recursion method}

\author{Jiaozi Wang~\orcid{0000-0001-6308-1950}}
\affiliation{Department of Mathematics/Computer Science/Physics, University of Osnabr\"uck, D-49076 
Osnabr\"uck, Germany}

\author{Mats H. Lamann}
\affiliation{Department of Mathematics/Computer Science/Physics, University of Osnabr\"uck, D-49076 
Osnabr\"uck, Germany}

\author{Robin Steinigeweg~\orcid{0000-0003-0608-0884}}
\affiliation{Department of Mathematics/Computer Science/Physics, University of Osnabr\"uck, D-49076 
Osnabr\"uck, Germany}
\author{Jochen Gemmer}
\affiliation{Department of Mathematics/Computer Science/Physics, University of Osnabr\"uck, D-49076 
Osnabr\"uck, Germany}

\date{\today}

\begin{abstract}
Understanding the transport behavior of quantum many-body systems constitutes
an important physical endeavor, both experimentally and theoretically. While
a reliable classification into normal and anomalous dynamics is known
to be notoriously difficult for a given microscopic model, even the seemingly
simpler evaluation of transport coefficients in diffusive systems continues to
be a hard task in practice. This fact has motivated the development and
application of various sophisticated methods and is also the main issue of
this paper. We particularly take a barely used strategy, which is based on the recursion method,
and demonstrate that this strategy allows for the  accurate
calculation of diffusion constants for different paradigmatic examples,
including magnetization transport in nonintegrable spin-$1/2$ chains and
ladders as well as energy transport in the mixed-field Ising model in one
dimension.
\end{abstract}

\maketitle

{\it Introduction.} Dynamics in quantum many-body systems constitutes a central
question in various disciplines of modern physics \cite{Bloch08, Polkovnikov11,
Lagen15, Gogolin15, Rigol16, Abanin19, RMP-transport}. In particular, transport
is a paradigmatic example of a dynamical process and is concerned with the
flow of a globally conserved quantity through a given system in the course
of time. While the study of transport has a long and fertile history, it 
continues to be in the focus of ongoing research \cite{RMP-transport}, both
experimentally and theoretically. A particular challenge is understanding the
behavior in the hydrodynamic regime, i.e., at large length scales and in the
long-time limit.

Within the class of physically relevant models, integrable systems play a
special role \cite{Caux_2011}. While the flow of energy is typically ballistic
in such systems, a richer phase diagram can result for other transport
quantities. A prime example in this context is the prominent spin-1/2 XXZ
model, which features ballistic, superdiffusive, and diffusive dynamics in
different parameter regions \cite{RMP-transport}. On the one hand, the origin of
ballistic dynamics has been traced back to (quasi-)local conserved quantities
\cite{Zotos97, Prosen2011, Prosen2013}. On the other hand, other non-ballistic
types of dynamics have been identified by the combination of complementary
analytical and numerical techniques over many years. Recent progress is based
on generalized hydrodynamics \cite{Bastianello2022, Doyon2023}, which yields a
comprehensive framework in all parameter regimes. Similarly, rich features can
be also found in quantum many-body models with disorder \cite{Nandkishore2014,
Luitz2017}, long-range interactions \cite{PhysRevA.90.063622-longrange,
PhysRevA.99.032114-longrange}, or constraints
\cite{PhysRevB.105.205127-constraint, PhysRevB.108.L020304-constraint} .

Integrable systems are the exception rather than the rule, and nonintegrable
systems represent the generic case. These systems add another level of
complexity, as they do not allow for an exact analytical treatment in the
thermodynamic limit. Still, a natural assumption for these models is the
emergence of diffusive dynamics \cite{RMP-transport}, e.g., due to the onset of
chaos and thermalization. However, this assumption has as such no consequence
for the value of the actual transport coefficient. And in fact, the intricate
calculation of transport in nonintegrable and integrable systems has been one
motivation for the development and application of sophisticated approaches,
where each approach has its own advantage and disadvantage. These approaches
range from perturbation theory \cite{Jung2006, Jung2007, Steinigeweg2016}, over
exact and Lanczos diagonalization \cite{HeidrichMeisner2007, Long2003,
Prelovsek2004}, quantum typicality \cite{Jin2021, Heitmann2020}, time-dependent
density-matrix renormalization group for correlation functions or wave packets
\cite{Langer2009, Karrasch2012, Karrasch2014}, quantum Monte-Carlo
\cite{Alvarez2002, Grossjohann2010}, Lindblad formulation of steady-state
transport \cite{Prosen2009, Znidaric2011, Wichterich2007}, to semi-classical and
classical treatment \cite{Wurtz2020, Schubert2021, McRoberts2022}, and probably
many more.

In view of this situation, this paper takes a fresh perspective and barely
explored strategy, which is based on the recursion method
\cite{viswanath1994recursion,Gray86,Parker19,Tolya20,PhysRevB.109.035117}.
{In one of the few works that pursue a similar approach, the
authors employ a so-called modified Mori theory to arrive at transport
coefficients \cite{Gray86}. (Results similar to those in Ref. \cite{Gray86} have
also been reported later in Ref.\ \cite{Lee01}). This approach requires the
knowledge of sufficiently many ``Lanczos coefficients" (for definition see
below). However, it was never applied to transport in specific interacting
quantum models, most likely since the computation of the Lanczos coefficients
practically requires modern computing power. Another way to compute  diffusion
constants from a finite set of Lanczos coefficients has recently been
suggested and applied to energy transport in a tilted-field Ising chain in Ref.\
\cite{Parker19}. This approach is based on the  {\it operator growth
hypothesis}, similar to the technique suggested in the current paper. The main
difference is that the latter employs the linear response theory, whereas the
former does not. As a consequence, the present approach is substantially more
efficient, especially in the thermodynamic limit. This opens for  otherwise
challenging analyses: For example, the insensitivity of the computed diffusion
coefficient to the number of  actually employed Lanczos coefficients, the
absence of which may indicate ballistic behavior \cite{supp}, becomes feasible, as will be
detailed below.}

We demonstrate that our strategy practically and consistently allows for the
accurate calculation of diffusion constants for different paradigmatic
examples, including magnetization transport in nonintegrable spin-1/2 chains
and ladders as well as energy transport in the mixed-field Ising model in one
dimension, and compare to values from the literature \cite{Robin14, Kloss2018,
Pollmann22, thomas2023comparing, artiaco2023efficient, Jonas19}.

{\it Models and Currents.} In this paper, we consider three different
paradigmatic examples of quantum many-body systems, which have also attracted
significant attention in the literature on transport before.
The first model is a perturbed XXZ spin-$1/2$ chain \cite{RMP-transport},
\begin{equation}
H = \sum_{r=1}^{L} (s_{r}^{x} s_{r+1}^{x} + s_{r}^{y} s_{r+1}^{y} + \Delta
s_{r}^{z}s_{r+1}^{z} + \Delta^\prime s_{r}^{z}s_{r+2}^{z} ) \,
,
\end{equation}
where $s_r^i$ ($i = x,y,z$) are the components of a spin-$1/2$ operator at
lattice site $r$, $L$ is the total number of lattice sites, and $\Delta$, $\Delta^\prime$ are anisotropies in
the $z$ direction. For $\Delta^\prime = 0$, the model is integrable in terms of
the Bethe Ansatz while, for any $\Delta^\prime \neq 0$, this integrability is
broken. For all values of $\Delta$, $\Delta^\prime$, the total magnetization
$S^z = \sum_{r} s_r^z$ is conserved, i.e., transport of local
magnetizations $s_r^z$ is a meaningful question. Note that, for this model and
the other models below, periodic boundary conditions are employed.

The second model is a XX spin-$1/2$ ladder \cite{Robin14, Kloss2018,
Pollmann22}. It is chosen as an
example for a quasi-onedimensional system, and its Hamiltonian is given by
$H = J_{\parallel}H_{\parallel}+J_{\perp}H_{\perp}$, where
\begin{eqnarray}
&& H_{\parallel} = \sum_{i=1}^{L} \sum_{l=1}^{2} (s_{r,l}^{x} s_{r+1,l}^{x} +
s_{r,l}^{y} s_{r+1,l}^{y}) \, , \nonumber \\
&& H_{\perp} = \sum_{r=1}^{L}(s_{r,1}^{x}s_{r,2}^{x}+s_{r,1}^{y}s_{r,2}^{y})
\end{eqnarray}
are the Hamiltonians of the legs and rungs, respectively, with corresponding
exchange coupling constants $J_\parallel$ and $J_\perp$. For $J_\perp
= 0$, the model is integrable and identical to noninteracting fermions by means
of the Jordan-Wigner transformation while, for any $J_\perp > 0$, the
integrability is broken again. As before, for all values of
$J_\parallel$, $J_\perp$, the total magnetization $S^z = \sum_{r,l} s_{r,l}^z$
is conserved.

The third and last model is the Ising spin-$1/2$ chain in the presence of a
mixed field \cite{thomas2023comparing, Pollmann22, artiaco2023efficient}. Its
Hamiltonian can be written as $H = \sum_{r=1}^L h_r$,
\begin{equation}
h_{r}=4s_{r}^{z}s_{r+1}^{z}+B_{x}(s_{r}^{x}+s_{r+1}^{x})+B_{z}(s_{r}^{z}+s_{r+1}^{z})\,,
\end{equation}
where $B_x$ and $B_z$ are the strengths of the field in $x$ and $z$ direction,
respectively. The integrability of the model for $B_x = 0$ or $B_z = 0$ is
broken for other values of $B_x$, $B_z$. In contrast to the previous models,
the total energy $H$ is the only non-trivial conserved quantity, i.e., transport of local
energies $h_r$ is meaningful here.

A natural strategy for the investigation of transport is given by currents
\cite{RMP-transport}.
The definition of local currents follows from the continuity equation
\begin{equation}
\dot{q}_r = i [H,q_r] = j_{r-1} - j_r \, ,
\end{equation}
where $q_r$ is the local transport quantity, i.e., either local magnetizations
$s_r$, $s_{r,l}$ or local energies $h_r$ for the models in this paper. Within
the theory of linear response, the autocorrelation function of the total
current $J = \sum_r j_r$ plays a central role and reads
\begin{equation} \label{ACF}
\langle J(t) J \rangle = \frac{\text{tr} [e^{-\beta H} e^{i H t} J e^{-i H t}
J]}{Z} \, , \quad Z = \text{tr} [e^{-\beta H}] \, ,
\end{equation}
where $\beta = 1/T$ is the inverse temperature, which we set to the still
nontrivial value $\beta = 0$ in the following. To
determine transport coefficients, one can then define the quantity
\cite{Robin09XXZ}
\begin{equation} \label{D}
D(t) = \frac{1}{\chi} \int_{0}^{t} \text{d}t' \, \langle J(t') J \rangle \, ,
\quad \chi = \langle Q^2 \rangle - \langle Q \rangle^2 \, ,
\end{equation}
where $Q = \sum_r q_r$ and $\chi$ is the static susceptibility. (For the three
models in this paper, expressions for $\chi$ and $J$ can be found in the
supplemental material \cite{supp}.) For a diffusive system, the diffusion constant is
given by $D = \lim_{t \to \infty}D(t)$, provided that the thermodynamic limit is
taken first. It is worth pointing out that the quantity $D(t)$ additionally
contains
information at finite time and length scales, as well as on other transport
types \cite{Robin09XXZ}. As apparent from Eq.\ (\ref{D}), it is crucial to
determine the area under the autocorrelation function. While various approaches
to this area have been applied in the literature before, we take in this paper a
barely used strategy, which is based on the recursion method.

{\it Framework.} In the following, it is quite convenient to switch to the
Hilbert of space of operators and denote its elements $O$ by states $|O)$. This
space is equipped with an inner product $(O_m|O_n) = \text{tr}[O^\dagger_m
O_n]$, which defines a norm via $|| O || = \sqrt{(O|O)}$. The Liouvillian
superoperator is defined by ${\cal L}|O) = [H, O]$ and
propagates a state $|O)$ in time, such that an autocorrelation function can be
written as $\langle O(t) O \rangle \propto C(t) = (O|e^{i{\cal L}t}|O)/|| O
||^2$.

The Lanczos algorithm can be employed to obtain a tridiagonal representation of
$\cal L$ in a subspace determined by some ``seed" $O$. In this paper, we have
$O = J$. To start the iterative scheme, we take a normalized initial state
$|O_0) \propto |O)$, i.e., $(O_0|O_0)=1$, and set $b_1 = \Vert {\cal L}O_0
\Vert$ as well as $|O_1) = {\cal L} |O_0)/b_1$. Then, we iteratively compute
\begin{eqnarray}
&& |O^\prime_n) = {\cal L} |O_{n-1}) - b_{n-1} |O_{n-2}) \, , \nonumber \\
&& b_n = || O^\prime_n || \, , \nonumber \\
&& |O_n) = |O^\prime_n)/b_n \, .
\end{eqnarray}
The tridiagonal representation of $\cal L$ in the Krylov basis
$\{ |O_n) \}$ results as
\begin{equation}
{\cal L}_{mn} = (O_{m}| {\cal L} |O_{n}) = \left(\begin{array}{cccc}
0 & b_{1} & 0 & \cdots \\
b_{1} & 0 & b_{2} \\
0 & b_{2} & 0 & \ddots \\
\vdots & & \ddots & \ddots
\end{array} \right)_{mn},
\end{equation}
where the Lanczos coefficients $b_n$ are real and positive numbers. Their
iterative computation is an elementary part of the recursion method. For the
remainder of this paper, we denote $|O_n)$ by $|n)$ for simplicity.

\begin{figure}[t!]
\includegraphics[width=0.85\linewidth]{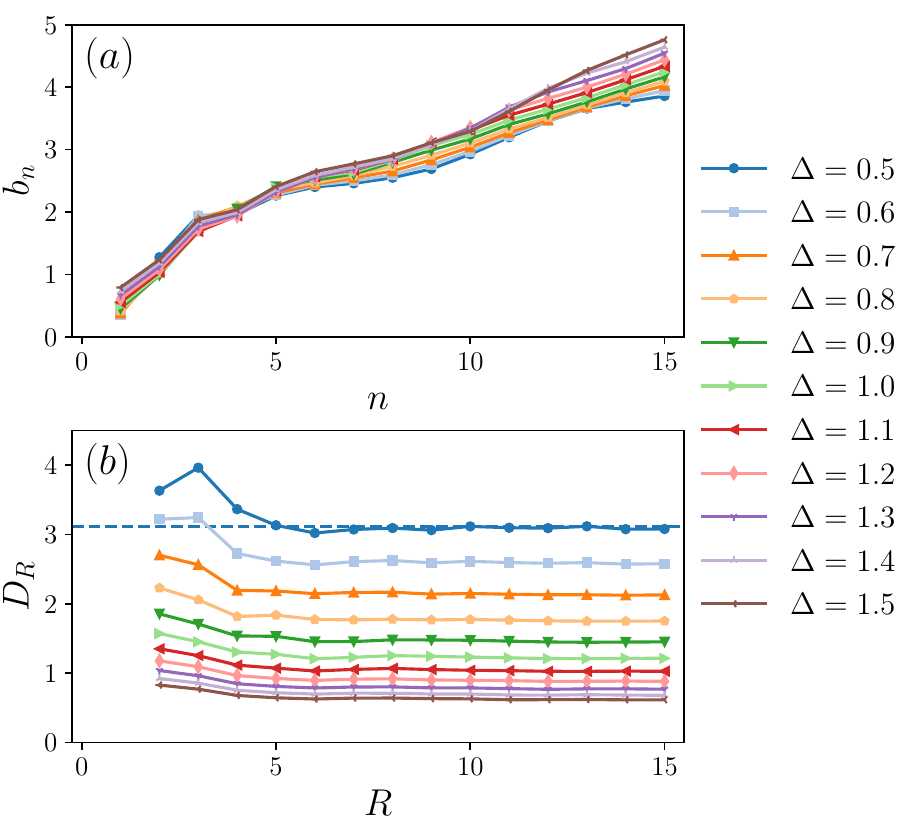}
\caption{Magnetization transport in the XXZ spin-$1/2$ chain with perturbation
$\Delta^\prime = 0.5$. (a) Lanczos coefficient $b_n$ vs.\ $n$ for
various $\Delta$. (b) Corresponding diffusion constants $D_R$. The dashed line
indicates $D = 3.1$ from Ref.\ \cite{Jonas19} for $\Delta =
0.5$.}
\label{Fig-Bn-XXZ}
\end{figure}

In the context of the Mori theory \cite{Gray86,10.1143/PTP.33.423-Mori}, the
time evolution of a set of functions $C_n(t)$ can be expressed in terms of a set
of integro-differential equations (see supplemental material for details),
\begin{equation} \label{eq-mc}
\dot{C}_{n}(t) = -b_{n+1}^{2} \int_{0}^{t} \text{d}t^{\prime} \,
C_{n+1}(t-t^{\prime}) \, C_{n}(t^{\prime}) \, ,
\end{equation}
$C_0(t) = C(t)$, and $C_{n}(t) = (n|e^{i{\cal L}_n t}|n)$. The operator ${\cal
L}_n$ is a ``submatrix" of ${\cal L}$ and defined as
\begin{gather}
{\cal L}_{n} = \sum_{m=n+1} b_{m} \Big[ |m+1)(m| + |m)(m+1| \Big] \, .
\end{gather}
Employing the Laplace transform of $C_n(t)$,
\begin{equation}
F_{n}(s) = \int_{0}^{\infty} \text{d}t \, e^{-st} \, C_{n}(t) \, ,
\end{equation}
one gets from Eq.\ (\ref{eq-mc}) the expression
\begin{equation}
F_{n}(s)=\frac{1}{s+b_{n+1}^{2}F_{n+1}(s)} \, .
\end{equation}
Thus, the Laplace transform of $C_0(t)$ can be written in the
form
\begin{equation} \label{eq-C0s}
F_{0}(s) = \frac{1}{s + \frac{b_{1}^{2}}{s + \frac{b_{2}^{2}}{s +
\frac{b_{3}^{2}}{\cdots}}}}.
\end{equation}
Then, recalling the definition of the diffusion constant in Eq.\ (\ref{D}),
one has $D = \langle J^2 \rangle F_{0}(0) / \chi$ and, using Eq.\
\eqref{eq-C0s} and iterating up to $R$, one has
\begin{equation}\label{eq-C0}
F_{0}(0) = \begin{cases}
F_{R}(0) \prod_{m=1}^{ \frac{R}{2}} \frac{b_{2m}^{2}}{b_{2m-1}^{2} } &
\text{, even } R\\
\frac{1}{b_{R}^{2} F_{R}(0)} \prod_{m=1}^{ \frac{R-1}{2} }
\frac{b_{2m}^{2}}{b_{2m-1}^{2}} & \text{, odd }R
\end{cases}.
\end{equation}
To further simplify the expression, we write ${F}_{R}(0)$ as
\begin{equation}
{F}_{R}(0)=\frac{1}{b_{R+1}}\int_{0}^{\infty}(R|e^{i\frac{{\cal L}_{R}}{b_{R+1}}(b_{R+1}t)}|R)\,\text{d}(b_{R+1}t)
\end{equation}
and ${F}_{R}(0) = p_{R+1}/b_{R+1}$
with the dimensionless quantity
\begin{equation}\label{eq-qn}
\!\!\!\!p_{R+1}\!=\!\int_{0}^{\infty}(R|e^{i\frac{{\cal L}_{R}}{b_{R+1}}t}|R)\,\text{d}t=\!\!\prod_{m=1}^{\infty}\!\!\left(\frac{b_{R+m}}{b_{R+m+1}}\right)^{(-1)^{m}}\!\!\!\!.
\end{equation}
Therefore, Eq.\ (\ref{eq-C0}) can be written as
\begin{equation} \label{eq-C0-New}
{F}_{0}(0)=\begin{cases}
\frac{1}{p_{R}b_{R}}\prod_{m=1}^{\frac{R}{2}}\frac{b_{2m}^{2}}{b_{2m-1}^{2}} & \text{, even }R\\
\frac{p_{R}}{b_{R}}\prod_{m=1}^{\frac{R-1}{2}}\frac{b_{2m}^{2}}{b_{2m-1}^{2}} & \text{, odd }R
\end{cases}.
\end{equation}


\begin{figure}[t!]
\includegraphics[width=0.9\linewidth]{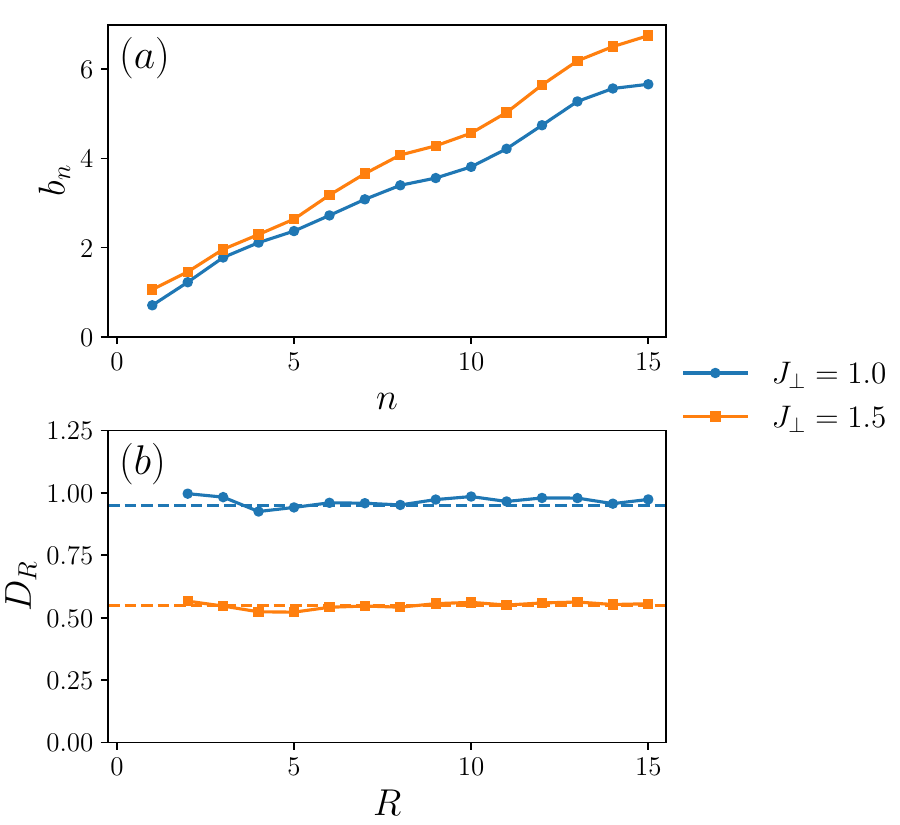}
\caption{Magnetization transport in the XX spin-$1/2$ ladder ($J_\parallel =
1$). (a) Lanczos coefficient $b_n$ vs.\ $n$ for $J_\perp = 1.0, 1.5$. (b)
Corresponding diffusion constants $D_R$. The dashed lines indicate
$D = 0.95$ (from Refs.\ \cite{Robin14, Kloss2018, Pollmann22}) and $D=0.55$
(from Ref.\ \cite{Robin14}) for $J_\perp = 1.0$ and $J_\perp = 1.5$,
respectively.}
\label{Fig-Bn-XXZ2L}
\end{figure}

Up to this point, everything is rigorous. If all $b_n$ are known, $p_R$ can be
calculated using Eq.\ (\ref{eq-qn}) and $F_0(0)$ can be obtained to arbitrary
precision. In practice, however, only several of the first $b_n$ are easily
accessible. So, the question is to find a good estimation of $p_R$ from those
$b_n$ which are numerically accessible. To this end, we employ the {\it operator
growth hypothesis} introduced in Ref.\ \cite{Parker19}, which states that, in a
chaotic system, the $b_n$ have for sufficiently large $n$  the asymptotic form
\begin{equation}\label{eq-ogh}
b_{n} = \begin{cases}
A \frac{n}{\ln n} + {\cal O}(\frac{n}{\ln n}) & , \, d=1 \\
\alpha n + \beta + {\cal O}(1) & \, , d > 1
\end{cases} \, .
\end{equation}
For dimension $d > 1$, neglecting the $ {\cal O}(1)$ term, one can derive an
analytical expression for $p_R$ \cite{supp},
\begin{equation}\label{eq-pn-origin}
p_{R}=\frac{\Gamma(\frac{R}{2}+\frac{\beta}{2\alpha})\Gamma(\frac{R}{2}+\frac{\beta}{2\alpha}+1)}{\Gamma^{2}(\frac{R}{2}+\frac{\beta}{2\alpha}+\frac{1}{2})}\ .
\end{equation}
For $d=1$, on top of the linear behavior, there is also a logarithmic
correction. If this correction only enters at very
large $n$, one can show that Eq.\ (\ref{eq-pn-origin}) still holds approximately \cite{supp}.

\begin{figure}[t!]
\includegraphics[width=0.9\linewidth]{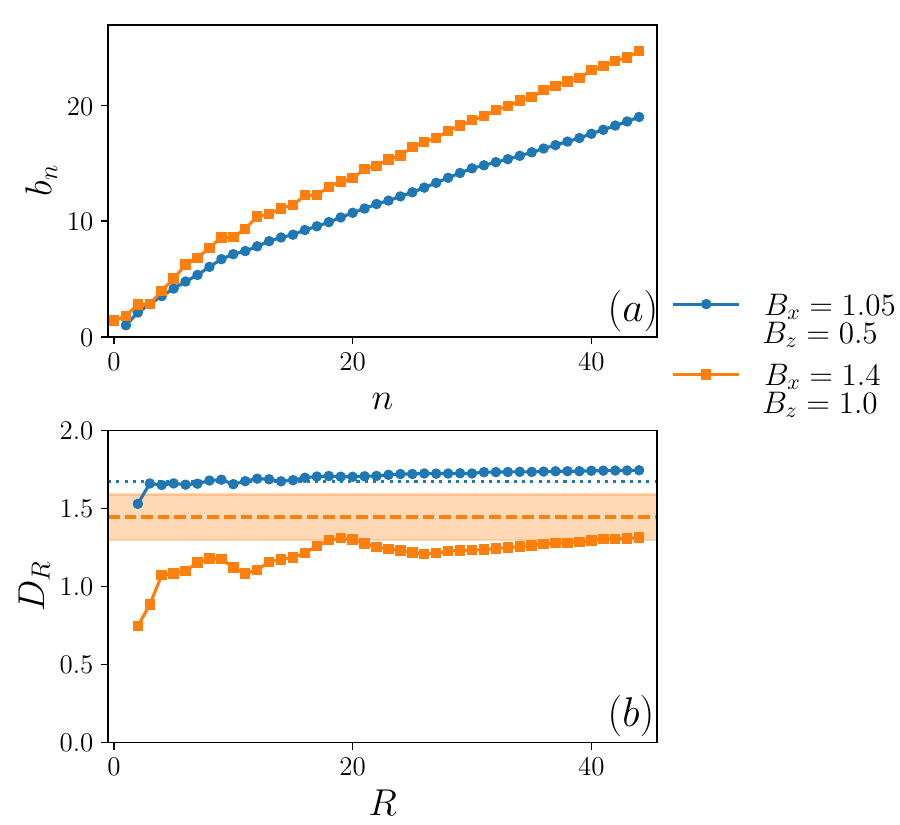}
\caption{Energy transport in the spin-$1/2$ Ising chain with a mixed field
($J = 1$). (a) Lanczos coefficient $b_n$ vs.\ $n$ for
various $B_z$. (b) Corresponding diffusion constants $D_R$. {The dotted line indicates $D=1.675$ \cite{Diff-DF} from Ref.\ \cite{Parker19} for $B_x = 1.05,\ B_z = 0.5$.} The dashed line
indicates $D = 1.44$ from Ref.\ \cite{thomas2023comparing} for
$B_x = 1.4,\ B_z = 0.9045$ (see also Refs.\ \cite{Pollmann22, artiaco2023efficient}), and our
result lies within a region of $10\%$ (shaded area).}
\label{Fig-Bn-Ising}
\end{figure}

Here, we use a rather simple approach, where $\alpha$ and $\beta$ are determined
by $b_R$ and $b_{R-1}$ only, i.e., $\alpha_R = b_{R} - b_{R-1}$ and
$\beta_R=Rb_{R-1}-(R-1)b_R$. This approach yields
\begin{equation} \label{eq-pn0}
p_{R}\simeq\tilde{p}_{R}=\frac{\Gamma(\frac{R}{2}+\frac{\beta_{R}}{2\alpha_{R}})\Gamma(\frac{R}{2}+\frac{\beta_{R}}{2\alpha_{R}}+1)}{\Gamma^{2}(\frac{R}{2}+\frac{\beta_{R}}{2\alpha_{R}}+\frac{1}{2})}\,.
\end{equation}
Then, substituting Eq.\ (\ref{eq-pn0}) to Eq.\ (\ref{eq-C0-New}), one obtains
an approximation of $F_0(0)$,
\begin{equation} \label{eq-mainresult}
F_{0}^{(R)}(0)=\begin{cases}
\frac{1}{\tilde{p}_{R}b_{R}}\prod_{m=1}^{\frac{R}{2}}\frac{b_{2m}^{2}}{b_{2m-1}^{2}} & \text{, even }R\\
\frac{\tilde{p}_{R}}{b_{R}}\prod_{m=1}^{\frac{R-1}{2}}\frac{b_{2m}^{2}}{b_{2m-1}^{2}} & \text{, odd }R
\end{cases}\,.
\end{equation}
It can be shown (in supplemental material \cite{supp}) that
$\tilde{p}_R\rightarrow 1$ for  $R \rightarrow \infty$ (or
$\frac{\beta_R}{\alpha_R} \rightarrow \infty$). 
In this large $R$ case Eq.\
\eqref{eq-mainresult} is almost identical to a result in Ref. \ \cite{Gray86}, which is based on an ad hoc assumption and has never been applied to specific many body systems, as already outlined in the introduction.
Correspondingly, one obtains an approximation of the diffusion constant,
\begin{equation} \label{eq-DQ-result}
D\simeq D_{R}=\frac{\langle J^{2}\rangle}{\chi}F_{0}^{(R)}(0)\,,
\end{equation}
which is a main result of this paper.

{\it Numerical Results.} Now, we check the estimation of $D$ in Eq.\
\eqref{eq-DQ-result} for the three different examples of quantum many-body
systems, with a focus on model parameters in the nonintegrable regime. First, we
numerically calculate the Lanczos coefficients $b_n$ in
Figs.\ \ref{Fig-Bn-XXZ} (a), \ref{Fig-Bn-XXZ2L} (a), and \ref{Fig-Bn-Ising} (a),
respectively. As visible in these figures, a region with an approximately
linear scaling of $b_n$ is observed in all cases considered.

Next, we depict the corresponding estimates of $D$ in Figs.\ \ref{Fig-Bn-XXZ}
(b), \ref{Fig-Bn-XXZ2L} (b), and \ref{Fig-Bn-Ising} (b). Remarkably, we observe
in Figs.\ \ref{Fig-Bn-XXZ} (b) and \ref{Fig-Bn-XXZ2L} (b) that $D_R$ saturates
at a constant value for $R \approx 5$ already, {while in Fig.\
\ref{Fig-Bn-Ising} (b) the saturation is less clear. 
To check whether the value of $D_R$ indeed provides a good estimate
of the true $D$,} we compare the results to existing numerical results in the
literature \cite{Robin14, Kloss2018, Pollmann22, thomas2023comparing,
artiaco2023efficient, Jonas19}, where values for $D$ have been determined for
the same model parameters. An almost perfect agreement with the estimate is
found in the XXZ model ($\Delta = 0.5$) and XX ladder shown in Figs.\
\ref{Fig-Bn-XXZ} (b) and \ref{Fig-Bn-XXZ2L} (b).  {
In the Ising model ($B_x=1.05,\ B_z=0.5$), $D_R (R\approx 40)$ approximately agrees with the results in Ref.~\cite{Parker19} with a deviation by a few percent \cite{Diff-DF}. 
Slightly larger deviation ($\approx 10\%$)
is observed at $B_x=1.4,\ B_z = 0.9045$ [Fig.\ \ref{Fig-Bn-Ising} (b)].}
Whether or not this deviation would vanish if more $b_n$ (beyond n=44) were
taken into account remains unclear within our investigation, due to numerical
limitations. However, up to $R\approx40$ the $D_R$ do not exhibit a systematic trend,
rather they vary visibly, thus indicating that an accurate result for the
diffusion constants might have not yet been reached. This is different from the
cases of the XXZ model and the XX ladder, where the $D_R$ as obtained from
Eq.\ \eqref{eq-DQ-result} converge very quickly to the corresponding results
from the literature.

Moreover, in the supplemental material \cite{supp}, we compare our results
in Figs.\ \ref{Fig-Bn-XXZ} (b), \ref{Fig-Bn-XXZ2L}
(b), and \ref{Fig-Bn-Ising} (b) to the exact calculation of $D$ in finite
systems, based on the relation in Eq.\ (\ref{D}). For the treatable system
sizes, we find convincing agreement, which supports the accuracy of the
estimate.

{\it Conclusion.} In summary, we have discussed in this paper an alternative
method for the accurate estimation of diffusion constants in quantum many-body
systems, which is based on the recursion method. By employing the {\it operator
growth hypothesis} in chaotic models, we have derived an estimate of the
diffusion constant. For several examples, we have found convincing agreement of
this estimate with results from the exact calculation in finite systems, and
with existing results in the literature. In particular, we have observed that
in many cases several of the first Lanczos coefficients are already sufficient
to get a good estimate, which can be obtained resource-efficiently in
comparison to other methods.

{\it Acknowledgements.} We thank Anatoly Dymarsky and Christopher White for
fruitful discussions. This work has been funded by the Deutsche
Forschungsgemeinschaft (DFG), under Grant No. 531128043, as well as under Grant
No.\ 397107022, No.\ 397067869, and No.\ 397082825 within the DFG Research
Unit FOR 2692, under Grant No.\ 355031190.

\bibliographystyle{apsrev4-1_titles}
\bibliography{Ref.bib}

\begin{thebibliography}{54}
\bibitem [{Dif()}]{Diff-DF}%
  \BibitemOpen
  \href@noop {} {}\bibinfo {note} {{Due to differences in the definition, the
  value of the diffusion constant reported in Fig.\ 7 of Ref.\ \cite{Parker19}
  differs from that in our paper by a factor of 2. More specifically, $D =
  3.35$ in Ref.\ \cite{Parker19} corresponds to $D = 1.675$ in ours. The
  parameters $J=1,\ B_x = 1.05,\ B_z = 0.5$ considered here is equivalent to
  the parameters considered in Fig.\ 7 of Ref.\ \cite{Parker19}}}\BibitemShut
  {NoStop}%
\end{thebibliography}

\clearpage
\newpage
\setcounter{figure}{0}
\setcounter{equation}{0}

\renewcommand{\thefigure}{S\arabic{figure}}
\renewcommand{\theequation}{S\arabic{equation}}
 
\section*{Supplemental material}
\subsection*{Derivation of Eq. (9)}\label{App::Addition}
In this section, we show the derivation of Eq. (9) in the main text. To this end, we define some ${\cal L}_n$ which is made from $\cal L$ by erasing the first $n$ row and the first $n$ column:
\begin{equation}
{\cal L}_{n}=\sum_{m=n+1}b_{m}\left(|m+1)(m|+|m)(m+1|\right).
\end{equation}
${\cal L}_n$ are Hermitian and we denote their eigenvectors
by $|k_n)$:
\begin{equation}
{\cal L}_n |k_n) = E_{k_n} |k_n).
\end{equation}

Firstly, we focus on ${\cal L}_1$. Making use of the basis $\{ |k_1), |0)  \}$, $\cal L$ can be written as
\begin{equation}
{\cal L}=\sum_{k_{1}} b_{1}\left\{|k_{1})(k_{1}|1)(0|+|0)(1|k_{1})(k_{1}|\right\}+E_{k_{1}}|k_{1})(k_{1}|.
\end{equation}
Now we switch to the interaction picture and taking ${\cal L}_1$ for the non-interacting system, the perturbation reads
\begin{equation}
{\cal V}_{I}(t)=\sum_{k_{1}} b_{1}\left\{|k_{1})e^{-iE_{k_{1}}t}(k_{1}|1)(0|+|0)(1|k_{1})e^{iE_{k_{1}}t}(k_{1}|\right\}.
\end{equation}
The time evolution in the interaction picture given by
\begin{equation}
i\frac{\partial}{\partial t}|\psi_{I}(t))={\cal V}_{I}(t)|\psi_{I}(t)),
\end{equation}
which in the basis $\{|k_1), |0) \}$ yields
\begin{gather}
\dot{r}_{0}=-ib_{1}\sum_{k_{1}}(1|k_{1})e^{iE_{k_{1}}t}r_{k_{1}} \nonumber \\
\dot{r}_{k_{1}}=-ib_{1}(k_{1}|1)e^{-iE_{k_{1}}t}r_{0},\label{eq-r0k}
\end{gather}
where 
\begin{equation}
r_{0}=(0|\psi_{I}(t)),\quad r_{k_{1}}=(k_{1}|\psi_{I}(t)).
\end{equation}
Integrating the second equation of Eq. \eqref{eq-r0k}, one obtains
\begin{equation}\label{eq-rk1}
r_{k_{1}}(t)=r_{k_{1}}(0)-ib_{1}(k_{1}|1)\int_{0}^{t}r_{0}(t^{\prime})e^{-iE_{k_{1}}t^{\prime}}dt^{\prime}.
\end{equation}
Choosing $r_{k_1}(0) = 0$ and inserting Eq. \eqref{eq-rk1} into the first line of Eq. \eqref{eq-r0k} yields
\begin{equation}
\dot{r}_{0}(t)=-b_{1}^{2}\int_{0}^{t}\sum_{k_{1}}|(k_{1}|1)|^{2}e^{iE_{k_{1}}(t-t^{\prime})}r_{0}(t^{\prime})dt^{\prime},
\end{equation}
which may be cast into the form of a standard integro-differential equations as
\begin{equation}
\dot{r}_{0}(t)=-\int_{0}^{t}K_{1}(t-t^{\prime})r_{0}(t^{\prime})dt^{\prime},
\end{equation}
where 
\begin{equation}
K_{1}(\tau)=\sum_{k_{1}}b_{1}^{2}|(k_{1}|1)|^{2}e^{iE_{k_{1}}\tau},
\end{equation}
The memory kernel $K_1(\tau)$ can also be written in a more compact form
\begin{equation}
K_{1}(\tau)=b_{1}^{2}C_1(\tau),
\end{equation}
where
\begin{equation}
C_1(\tau) = (1|e^{i{\cal L}_{1}\tau}|1).
\end{equation}
Noting that the probability at $|0)$ are the same in the interaction picture as in the Schr\"{o}dinger picture, one has
\begin{equation}
C_0(t) = r_0(t) = -b^2_1\int_0^t C_1(t-t^\prime) C_0 (t^\prime) dt^\prime.
\end{equation}
Repeating the above procedure, it is easy to get 
\begin{equation}
C_{n}(t)=-b_{n+1}^{2}\int_{0}^{t}C_{n+1}(t-t^{\prime})C_{n}(t^{\prime})dt^{\prime},
\end{equation}
where $C_n(\tau)$ is defined as
\begin{equation}
C_n(\tau) = (n|e^{i{\cal L}_n \tau}|n).
\end{equation}

\begin{figure}[t!]
	\centering
	\includegraphics[width = 0.85\linewidth]{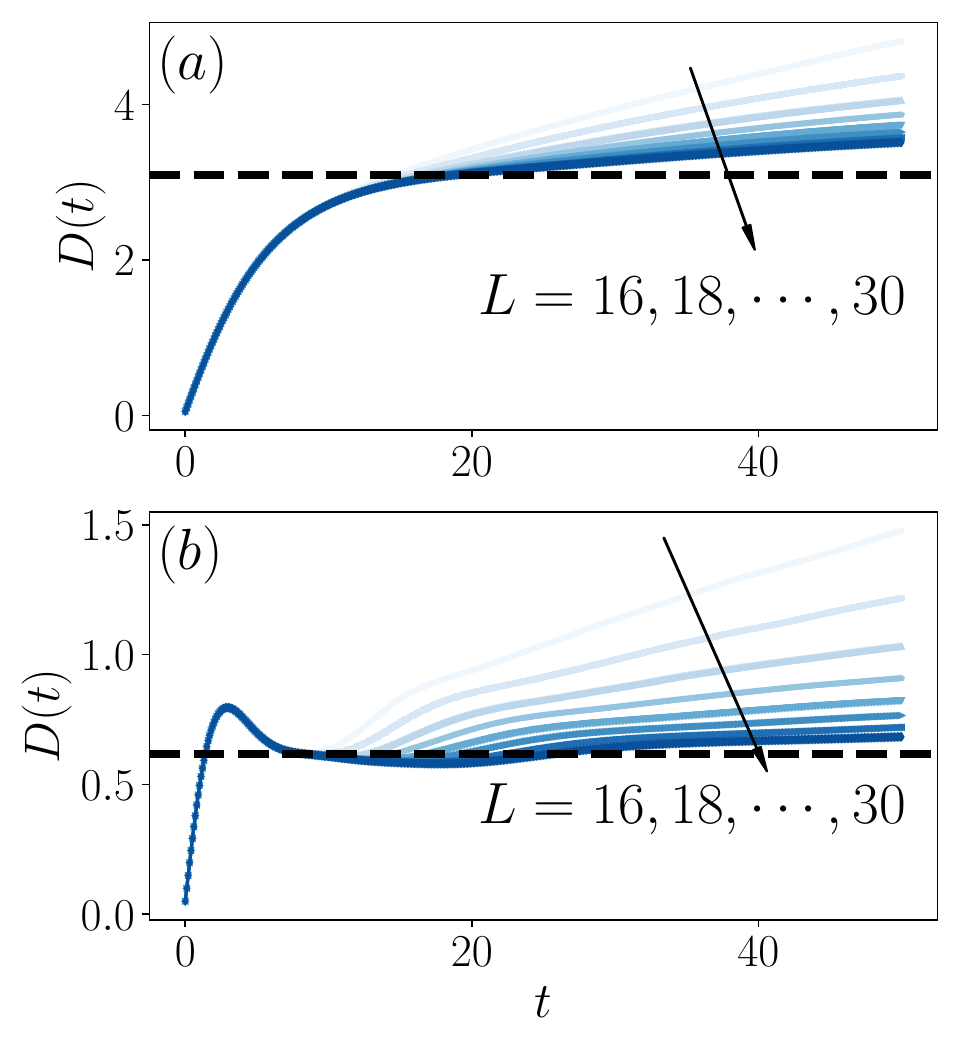}
	
	\caption{Time-dependent diffusion constant $D(t)$ in the XXZ chain with
		parameters $\Delta^\prime = 0.5$ and (a) $\Delta = 0.5$ and (b) $\Delta = 1.5$,
		for system size $L = 16,18,\cdots,30$ (from light to dark). The dashed line indicates the average
		value of the last five $D_R$ shown in Fig. 1(b). For (a), data
		for larger system sizes $L=33$ can also be found in Ref. \cite{Jonas19}.}
	\label{Fig-Dt-XXZ}
\end{figure}

\begin{figure}[t!]
	\includegraphics[width = 0.85\linewidth]{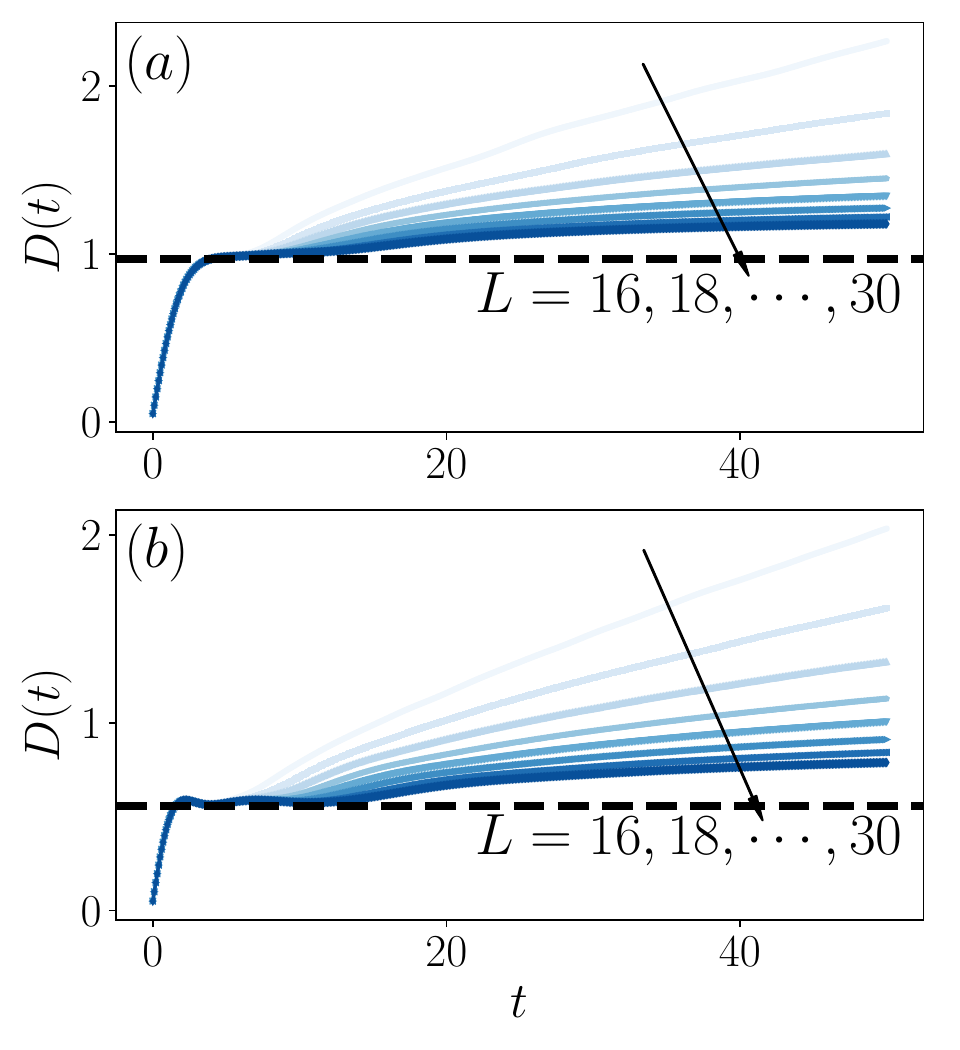}
	\caption{Time-dependent diffusion constant $D(t)$ in the XX ladder with
		parameters $J_{\parallel} = 1.0$ and (a) $J_\perp = 1.0$; (b) $J_\perp = 1.5$,
		for system size $L = 16,18,\cdots,30$ (from light to dark). The dashed line indicates the average
		value of the last five $D_R$ shown in Fig.\ 2 (b). Data for
		larger system sizes $L=34$ can also be found in Ref.\ \cite{Robin14}.}
	\label{Fig-Dt-XXZ2L}
\end{figure}

\begin{figure}[!]
	\includegraphics[width = 0.85\linewidth]{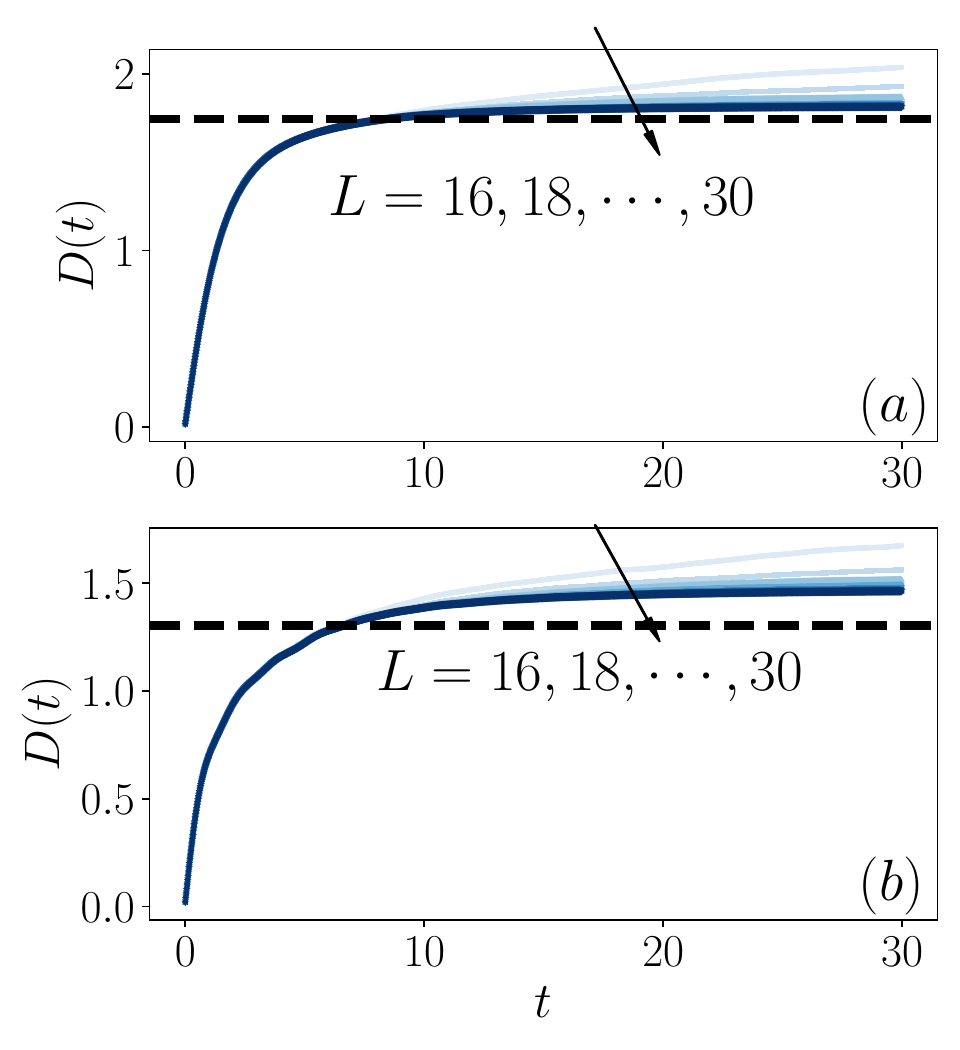}
	\caption{Time-dependent diffusion constant $D(t)$ in the mixed-field
		Ising chain with parameters (a) $B_z = 0.5,\ B_x = 1.05$ and (b) $B_z = 0.9045,\ B_x = 1.4$, 
		for system size $L = 16,18,\cdots,30$ (from light to dark). The dashed line indicates the average
		value of the last five $D_R$ shown in Fig. 3(b). Data for
		larger system sizes can also be found in Refs.\ \cite{thomas2023comparing,
			artiaco2023efficient, Pollmann22}.} \label{Fig-Dt-Ising}
\end{figure}

\subsection*{Derivations of Eq.\ (19)}
In this section, we show the detailed derivation of Eq. (19) in the main text.
Inserting $b_n = \alpha n + \beta$ to Eq.  (16) yields
\begin{gather}
p_{R}=\prod_{k=1}^{\infty}\frac{(b_{2k+R-1})^{2}}{b_{2k+R}b_{2k-2+R}}\nonumber\\
=\prod_{k=1}^{\infty}\frac{\left(\alpha(2k+R-1)+\beta\right)^{2}}{\left(\alpha(2k+R)+\beta\right)\left(\alpha(2k-2+R)+\beta\right)} \nonumber \\
=\prod_{k=1}^{\infty}\frac{\left(k+\frac{\alpha R+\beta}{2\alpha}-\frac{1}{2}\right)^{2}}{\left(k+\frac{\alpha R+\beta}{2\alpha}\right)\left(k+\frac{\alpha R+\beta}{2\alpha}-1\right)}. \label{eq-pn1}
\end{gather}
Making use the following expression of Gamma function 
\begin{equation}
\Gamma(z+1)=\prod_{k=1}^{\infty}\left[\frac{1}{1+\frac{z}{k}}\left(1+\frac{1}{k}\right)^{z}\right],
\end{equation} 
it is easy to get 
\begin{equation}\label{eq-GammaP}
\frac{\Gamma(z-a+1)\Gamma(z+a+1)}{\Gamma^{2}(z+1)}=\prod_{k=1}^{\infty}\frac{(k+z)^{2}}{(k+z-a)(k+z+a)}.
\end{equation}
Comparing Eq. \eqref{eq-GammaP} with Eq. \eqref{eq-pn1},
and setting $z=\frac{R}{2}+\frac{\beta}{2\alpha}-\frac{1}{2},\quad a = \frac{1}{2}$, one has
\begin{equation}
p_{R}=\frac{\Gamma(\frac{R}{2}+\frac{\beta}{2\alpha})\Gamma(\frac{R}{2}+\frac{\beta}{2\alpha}+1)}{\Gamma^{2}(\frac{R}{2}+\frac{\beta}{2\alpha}+\frac{1}{2})},
\end{equation}
which is the result of Eq. (19).

\subsection*{Logarithmic correction of $b_n$}
The derivation of our main result in Eq. (19) is based on a linear asymptotic form of Lanczos coefficient $b_n$.
But it is shown in Ref. \cite{Parker19} that on top of the main linear behavior, for 1d system, there is a logarithmic correction for large $n$.
In this section, we show that in this case, Eq. (19) still holds approximately.

\begin{figure}[t]
	\centering\includegraphics[width = 0.95\linewidth]{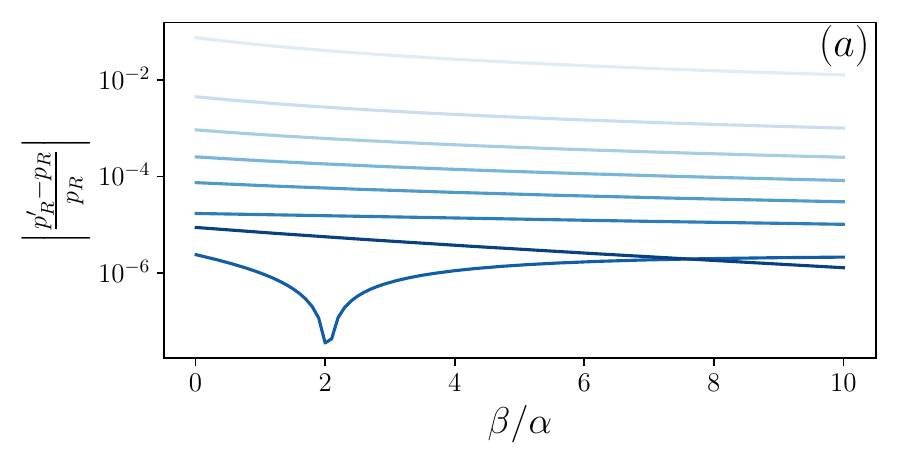}
	\bigskip\hrulefill\bigskip	
	\includegraphics[width = 0.9\linewidth]{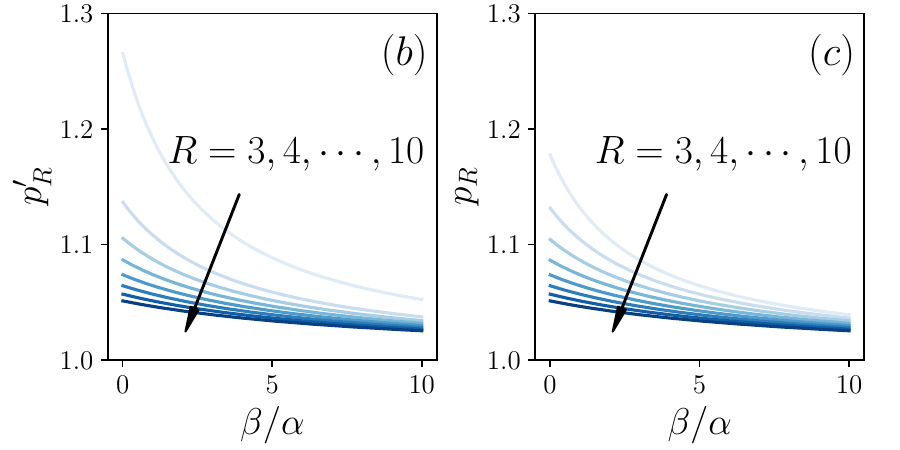}
	\caption{(a) $\left|\frac{{p}^\prime_{R}-p_{R}}{p_{R}}\right|$, (b) ${p}^\prime_R$ and (c) $p_R$ versus $\beta/\alpha$ for various $R=3,4,\cdots,10$ (from light to dark). }\label{Fig-Check}
\end{figure}

We consider $p_R$ and ${p}^\prime_R$ which are defined as
\begin{equation}\label{eq-pn2}
p_{R}=\prod_{k=1}^{\infty}\frac{(b_{2k+R-1})^{2}}{b_{2k+R}b_{2k-1+R}},\quad{p}^\prime_{R}=\prod_{k=1}^{\infty}\frac{({b}^\prime_{2k+R-1})^{2}}{{b}^\prime_{2k+R}{b}^\prime_{2k-1+R}},
\end{equation}
where
\begin{equation}\label{eq-bn2}
\begin{cases}
b_{n}=\alpha n+\beta\\
{b}^\prime_{n}=A\frac{n}{\log n}+B
\end{cases} \text{for}\ n\ge R.
\end{equation}
Here
\begin{equation}\label{eq-AB}
A=\frac{(\log R)^{2}}{\log R-1}\alpha,\  B=\beta-\frac{\alpha R}{\log R-1},
\end{equation}
are chosen such that the value of of $b_n$ and ${b}^\prime_n$  as well as their first derivative coincide at $n=R$. 
Inserting Eq.  \eqref{eq-bn2} to Eq. \eqref{eq-pn2} one gets
\begin{gather}
{p}^\prime_R=
\prod_{k=1}^{\infty}\frac{\left(\frac{2k+R-1}{\log(2k+R-1)}+\frac{B}{A}\right)^{2}}{\left(\frac{2k+R}{\log(2k+R)}+\frac{B}{A}\right)\left(\frac{2k+R-2}{\log(2k+R-2)}+\frac{B}{A}\right)},  \label{eq-prtilde}\\
{p}_{R}=\prod_{k=1}^{\infty}\frac{\left(2k+R-1+\frac{\beta}{\alpha}\right)^{2}}{\left(2k+R+\frac{\beta}{\alpha}\right)\left(2k+R-2+\frac{\beta}{\alpha}\right)}. \label{eq-pr}
\end{gather}
According to Eq. \eqref{eq-AB} one has
\begin{equation}\label{eq-AB0}
\frac{B}{A}=\frac{\beta}{\alpha}\frac{\log R-1}{(\log R)^{2}}-\frac{R}{(\log R)^{2}}.
\end{equation}
From Eqs. \eqref{eq-prtilde}--\eqref{eq-AB0}, one can see that ${p}^\prime_R$ and ${p}_R$ depend only on $R$ and $\beta/\alpha$.
Different from $p_R$, we do not have an analytical expression for
${p}^\prime_R$. But the convergence of the infinite product series in
${p}^\prime_R$ can be proved by making use of Leibniz criterion.
In numerical simulations, ${p}^\prime_R$ is estimated by keeping the product series to $k=5\times10^6$.
In Fig. \ref{Fig-Check} (a) we plot the difference between ${p}^\prime_R$ and ${p}_R$ as a function of $\frac{\beta}{\alpha}$ for various $R$. It is observed that for $R\ge 4$, the relative difference $\left|\frac{{p}^\prime_{R}-p_{R}}{p_{R}}\right|$ is below $1\%$ for all values of $\beta/\alpha$ we consider. It indicates that Eq. (19) still holds approximately in presence of the logarithmic correction. In addition, in Fig. \ref{Fig-Check} (b) and (c), we show $p^\prime_R$ and $p_R$, from which one can see that both $p^\prime_R$ and $p_R$ goes to $1$ if either $R$ or $\beta/\alpha$ goes to infinity.  
\ \
\\

\subsection*{Diffusion constant in integrable models}
\begin{figure}[t]
\centering\includegraphics[width = 0.95\linewidth]{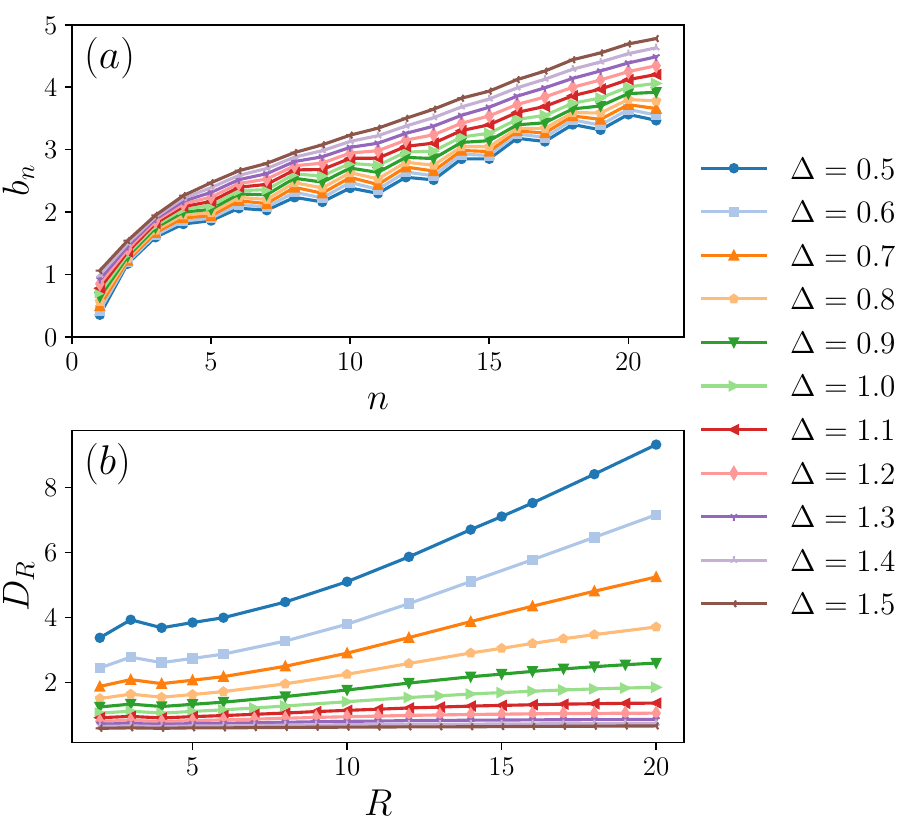}
\caption{Magnetization transport in the integrable XXZ spin-$1/2$ chain
[defined in Eq.~\eqref{eq-XXZ-Int}]. (a) Lanczos coefficient $b_n$ vs.\ $n$ for
various $\Delta$. (b) Corresponding diffusion constants $D_R$.}\label{Fig-XXZ-Int}
\end{figure}

{In addition to the non-integrable models studied in the main
text, we also consider an integrable model, i.e., a spin-$1/2$ XXZ chain with
only nearest-neighbour coupling,
\begin{equation}\label{eq-XXZ-Int}
H = \sum_{r=1}^{L} (s_{r}^{x} s_{r+1}^{x} + s_{r}^{y} s_{r+1}^{y} + \Delta
s_{r}^{z}s_{r+1}^{z}) \ .
\end{equation}
It is well-known that magnetization transport depends on the value of
$\Delta$ \cite{RMP-transport}: It is ballistic at $\Delta < 1$, normal
diffusive at $\Delta > 1$, and super-diffusive at $\Delta = 1$.
In Fig.~\ref{Fig-XXZ-Int}, we show the estimation of the diffusion constant by
Eq.\ (22) in the main text. A substantial difference in the behavior of $D_R$
is observed for $\Delta < 1$ compared to $\Delta >1$: $D_R$ appears to diverge
for $\Delta < 1$ and tends to converge to some finite value for $\Delta > 1$.
The (seemingly) divergence of $D_R$ is due to the pronounced even-odd effect,
in addition to the dominant linear increase of Lanczos coefficients, i.e.,
\begin{equation}
b_{n}\sim\alpha n+\beta+(-1)^{n}\gamma_{n} \, .
\end{equation}
Here, we are not aiming at  a precise location of the transition from ballistic
to diffusive transport by making use of the recursion method. This  would
definitely need the knowledge of infinite number of Lanczos coefficient, which
is clearly beyond the reach of the numerical method at hand. The primary
objective of this example is to show that ``finite size scaling'' of $D_R$
(dependence of $D_R$ on $R$) can at least show some hint on the onset of the
ballistic (or at least super-diffusive) behavior, particularly when an overall
increase of $D_R$ is observed.}

\subsection*{Time-dependent diffusion constant in finite-size systems}
In the spin-$1/2$ XXZ chain and XX ladder, we consider the spin current
\begin{equation}
J_S =\begin{cases}
 \sum_{r=1}^{L}(s_{r}^{x}s_{r+1}^{y}-s_{r}^{y}s_{r+1}^{x}), & \text{XXZ}\\
J_{\parallel}\sum_{r=1}^{L/2}\sum_{k=1}^{2}(s_{r,k}^{x}s_{r+1,k}^{y}-s_{r,k}^{y}
s_{r+1,k}^{x}), & \text{XX}
\end{cases}
\end{equation}
with the time-dependent diffusion coefficient
\begin{equation}
D(t) = \frac{1}{\chi}\int_{0}^{t} \langle J_{S}(t')
J_{S} \rangle \, dt' \, , \quad \chi = \frac{L}{4} \, .
\end{equation}
In the mixed-field Ising model, we instead consider the energy-current
operator
\begin{equation}
J_{E}=4B_{x}\sum_{r=1}^{L}s_{r}^{y}(s_{r+1}^{z}-s_{r-1}^{z})
\end{equation}
with the time-dependent diffusion constant
\begin{equation}
D(t)  =\frac{1}{\chi}\int_{0}^{t} \langle
J_{E}(t') J_{E}\rangle \, dt' \, , \quad \chi=
{L} (B^2_x+B^2_z+1) \, .
\end{equation}

In Figs. \ref{Fig-Dt-XXZ}, \ref{Fig-Dt-XXZ2L} and \ref{Fig-Dt-Ising}, we
calculate the time-dependent diffusion coefficient $D(t)$ for finite systems,
using dynamical typicality \cite{Jin2021, Heitmann2020}. In all three models,
we can see a tendency that the long-time value $D(t)$ becomes closer to our
estimation $\overline{D_R}$ (average over the last $5$ values) for
larger $L$. Whether or not this value will converge to $\overline{D_R}$ is in
principle not clear, due to the limited system size ($L \leq 30$).
But based on the results, it is reasonable to expect that the diffusion
constant $D$($L \rightarrow \infty$) is close to $\overline{D_R}$.

\end{document}